\documentstyle[preprint,aps]{revtex}
\begin{document}
\draft
\author{J. Dewitz, W. H\"ubner, and K. H. Bennemann}
\address{Institut f\"ur Theoretische Physik der Freien Universit\"at
Berlin,Arnimallee 14, 14195 Berlin, Germany.}
\date{September 30,1994}
\title{Theory for Nonlinear Mie-Scattering from Spherical Metal Clusters}
\maketitle

\begin{abstract}
Using classical electrodynamics we determine the angular dependence of the
light intensities radiated in second and third harmonic generation by spherical
 metal clusters. Forward and backward scattering is analyzed in detail. Also
resonance effects in the integrated intensities are studied. Our work treats
the case of intermediate cluster sizes. Thus it completes the scattering
theory fo spherical clusters between {\em Rayleigh}-type analysis for small
spheres and geometrical optics for spheres much larger than the wavelength for
 nonlinear optics. Since the particle size sensitivity of {\em Mie}-scattering
 is increased by nonlinearity, the results can be used to extract sizes of
small particles from nonlinear optics.
\end{abstract}
\newpage

\section{Introduction}
In this paper we investigate the nonlinear interaction of light with small
metal particles. In nonlinear optics one expects a more pronounced sensitivity
 of the radiated yield to the light-wavelength and cluster-size, due to
effects by nonlinear sources.

The linear interaction of light with spherical objects shows certain changes
depending on the size of the sphere. The limiting cases are {\em Rayleigh}-
scattering for spheres small compared with the wavelength and reflection by
spheres much larger than the wavelength (the limit of geometrical optics). If
the size of the sphere and the wavelength of light have comparable magnitudes
the intensities scattered by spherical metal clusters are strongly enhanced.
Mie \cite{Mie} first calculated this classical effect and found that most of
the incident light intensity is scattered in the forward direction, because of
 interference of waves originating from the front and the back of the sphere.
This so called {\em Mie-effect} is increased with increasing sphere size.
Bohren and Huffmann \cite{BH} pointed out that the extinction efficiency
$Q_{ext}$ defined as the sum of scattering and absorption efficiencies
($Q_{ext}=Q_{sc}+Q_{abs}$) as a function of the sphere-radius are
characterized by the interference- and ripple structure. The interference
structure gives rise to regular interferences of incident and forward
scattered light dominating the envelope of $Q_{ext}$, see Fig.
(\ref{linQsc133.0}). Superimposed on this is sharp and highly irregular
ripple structure originating from resonant electromagnetic surface modes of
the sphere. In general, the resonances depend on the size parameter $ka$
comparing the radius of the sphere $a$ and the wavelength expressed by the
absolute value of the wave vector $k$ of the incident light. Thus resonances
also occur by varying the wavelength of the incident light. For small spheres
only one peak or resonance appears at $\omega_p/\sqrt{3}$ with $\omega_p$ the
plasma frequency and the $1/\sqrt{3}$-factor caused by the shape of the
sphere. This is the dipole-term. The larger the size of the sphere the more
resonances around the $\omega_p/\sqrt{3}$-peak appear (if $ka>1$) reflecting
the surface-mode resonances or higher multipoles, see\cite{Martin}.

The nonlinear interaction of light with small particles was recently studied
by \"Ostling {\em et al.} \cite{Stampf} using a simplified model to obtain the
 intensities in second and third harmonic generation, SHG and THG,
respectively. The calculations yield an enhancement of the $total$ intensities
 by a factor 5000 in SHG and 200000 in THG in comparison to a plane metal
surface for spheres with sizes $a>\lambda_p/2$ in a wide frequency range.
Here, $\lambda_p$ is the wavelength corresponding to $\omega_p$ using the
relation $c=\lambda\cdot\omega/2\pi$ with $c$ the speed of light in vacuum.
This holds especially if the diameter of the sphere roughly equals $\lambda_p$.

The correlation between the scattering behavior and the properties of the
sphere enables us to obtain the size and complex refractive index of the
sphere and in the case of an ensemble of spheres even the size distribution
from the scattered intensities. It is a goal of our investigation to show that
 this is improved by higher harmonics.
In this paper we extend the theory of ref. \cite{Stampf} to obtain the {\em
angular dependence} of the radiated second and third harmonic intensities,
compare the linear and nonlinear results examining their characteristic
features in higher harmonics. In particular the angular resolved intensities,
the degree of polarization of the scattered SHG and THG yield are
investigated. In addition, we thoroughly study parameter dependences of the
ratio of forward and backward scattering and analyze the origin of  the
resonance structures.

Our main results are an enhanced size sensitivity of the higher harmonics
compared to linear $Mie$-scattering, thus pronouncing the mentioned
characteristic features of the linear {\em Mie-effect} in higher harmonic
scattering. Similarities between the linear scattering and THG reflect that
the scattering is dominated by the same multipoles in both cases. In contrast,
 different multipoles contribute to SHG resulting in distinct changes of the
angular dependence. Because of the more pronounced forward scattering enhanced
 backward scattering vanishes in the higher harmonics.

The following section II contains the appropriate theoretical calculations.
In section III we present the numerical results starting with polar plots of
the intensities in SHG and THG, plots for the degree of polarization, the
ratio of forward and backward scattering and the integrated intensities as a
function of the size of the spheres. In section IV we will give a summary of
the results.
\section{Theory}
First, we use Mie theory to calculate the linear scattering and the radial
component of the polarization $\sigma\left(\theta,\phi\right)$ at the surface
of the sphere. Then, we determine the radiated intensities in second and third
 harmonic by matching the electromagnetic fields and the $n$-th power of
$\sigma\left(\theta,\phi\right)$ with regard to the appropriate boundary
conditions, as described by \"Ostling {\em et al.} \cite{Stampf}. In section
II.D the characteristic quantities of the radiation are determined. These
quantities can be compared with experiments.
\subsection{Linear scattering by spheres}
According to the spherical symmetry of the problem we expand the fields in
the form
%
\begin{eqnarray}
{\bf E}^i\left({\bf x}\right)&=&\sum_{l,m}C\left(l\right)\left[a^i_M
\left(l,m\right)
f^i_l\left(k_1r\right){\bf X}_{l,m}\left(\theta,\phi\right)\right.\nonumber\\
& &\left.\qquad\qquad\qquad+\frac{m}{\left|m
\right|}a^i_E\left(l,m\right)\frac{1}{\epsilon\left(\omega\right)k}\nabla
\times f^i_l\left(k_1r\right){\bf X}_{l,m}\left(\theta,
\phi\right)\right]
\label{Elin}
\end{eqnarray}
Therein, ${\bf X}_{l,m}$ is a vector spherical harmonic as introduced by
Jackson \cite{Jackson} with $C\left(l\right)=i^l\sqrt{4\pi\left(2l+1\right)},
k=\omega/c$ and $k_1=\sqrt{\epsilon\left(\omega\right)}k$ . The multipole
coefficients $a_M^i\left(l,m\right)$ and $a_E^i\left(l,m\right)$ refer to the
magnetic (transverse electric) and electric (transverse magnetic) multipoles.
The index $i$ specifies the incident ($i\equiv inc$), the scattered ($i\equiv
sc$) or the internal ($i\equiv in$) fields. For incident waves of positive and
 negative helicity we have $a_M^{inc}\left(l,\pm1\right)=a_E^{inc}\left(l,
\pm1\right)=1$. In this paper we use a superposition of both to give linear
polarization. The spherical Hankel functions
$f^{sc}_l\left(kr\right)=h_l^{\left(
1\right)}\left(kr\right)$ and Bessel functions $f^{inc,in}_l\left(kr\right)=j_
l\left(kr\right)$ describe the radial part of the field outside and inside
the sphere. The magnetic field is given by the Maxwell-equation for harmonic
fields
\begin{equation}
{\bf B}=-i\omega/c\cdot{\bf\nabla}\times{\bf E}\nonumber.
\end{equation}
Using the boundary conditions at the surface of the sphere\\
%
\begin{equation}
{\bf n}\times\left({\bf E}^{sc}+{\bf E}^{inc}\right)={\bf n}\times{\bf E}^
{in}
\label{GrenzbedlinE}
\end{equation}
and
%
\begin{equation}
{\bf n}\times\left({\bf B}^{sc}+{\bf B}^{inc}\right)={\bf n}\times{\bf B}^
{in},
\label{GrenzbedlinB}
\end{equation}
we obtain the expansion coefficients of the scattered wave
%
\begin{eqnarray}
a_E^{sc}\left(l,\pm1\right)&=&\left.\frac{j_l\left(kr\right)\frac{\partial}
{\partial r}\left[rj_l\left(k_1r\right)\right]-\epsilon\left(\omega\right)
j_l\left(k_1r\right)\frac{\partial}{\partial r}\left[rj_l\left(kr\right)
\right]}{\epsilon\left(\omega\right)j_l\left(k_1r\right)\frac{\partial}
{\partial r}\left[rh_l^{\left(1\right)}\left(kr\right)\right]-h_l^{\left(1
\right)}\left(kr\right)\frac{\partial}{\partial r}\left[rj_l\left(k_1r\right)
\right]}\right|_{r=a},\nonumber\\
a_M^{sc}\left(l,\pm1\right)&=&\left.\frac{j_l\left(kr\right)\frac{\partial}
{\partial r}\left[rj_l\left(k_1r\right)\right]-
j_l\left(k_1r\right)\frac{\partial}{\partial r}\left[rj_l\left(kr\right)
\right]}{j_l\left(k_1r\right)\frac{\partial}
{\partial r}\left[rh_l^{\left(1\right)}\left(kr\right)\right]-h_l^{\left(1
\right)}\left(kr\right)\frac{\partial}{\partial r}\left[rj_l\left(k_1r\right)
\right]}\right|_{r=a}
\label{linkoeff}
\end{eqnarray}
at the surface of the sphere with radius $a$. From the continuity of the
electrical displacement at the surface of a perfect conductor
%
\begin{equation}
{\bf n}\cdot\left({\bf D}^{sc}+{\bf D}^{inc}\right)={\bf n}\cdot{\bf D}^{in},
\label{GrenzbedMetall}
\end{equation}
the surface charge results as
%
\begin{equation}
\sigma\left(\theta,\phi\right)=\frac{1}{4\pi}Re\left[\left({\bf E}^{sc}+
{\bf E}^{inc}-{\bf E}^{in}\right)\cdot{\bf n}\right]e^{-i\omega t},
\label{sigma}
\end{equation}
where ${\bf n}={\bf r}/\left|{\bf r}\right|$ and $Re$ denots the real part.
Furthermore we expand $\sigma
\left(\theta,\phi\right)$ in spherical harmonics
%
\begin{equation}
\sigma\left(\theta,\phi\right)=\frac{1}{2}\sum_{l,m=\pm1}a_{l,m}^{\left(1
\right)}Y_{l,m}\left(\theta,\phi\right)e^{-i\omega t}+c.c.\quad.
\label{sigmaent}
\end{equation}
The expansion coefficients result from the orthogonality of the spherical
harmonics as
%
\begin{equation}
a_{l,\pm1}^{\left(1\right)}=\frac{1}{4\pi}\left(1-\frac{1}{\epsilon\left(
\omega\right)}\right)\frac{C\left(l\right)i\sqrt{l\left(l+1\right)}}{ka}
\left(j_l\left(ka\right)+a_E^{sc}\left(l,\pm1\right)h_l^{\left(1\right)}\left(
ka\right)\right).
\label{al1}
\end{equation}
%
\subsection{Sources of the higher harmonic radiation}
In analogy to the linear case we expand the n-th power of the surface charge
$\sigma$ in terms of spherical harmonics:
%
\begin{equation}
\sigma^n\left(\theta,\phi\right)=\frac{1}{2}\sum_{l,m}a_{l,m}^{\left(n\right)}
Y_{l,m}\left(\theta,\phi\right)e^{-ni\omega t}+c.c.\quad .
\label{sigmahochn}
\end{equation}
Neglecting time-independent terms we obtain the coefficients in the case of
second harmonic generation as
%
\begin{eqnarray}
a_{l,2}^{\left(2\right)}&=&\frac{1}{2}\sum_{l_1=1}^{\infty}\sum_{l_2=1}^{
\infty}a_{l_1,1}^{\left(1\right)}a_{l_2,1}^{\left(1\right)}\int Y_{l,2}^*Y_{
l_1,1}
Y_{l_2,1}d\Omega\;,\nonumber\\
a_{l,-2}^{\left(2\right)}&=&\frac{1}{2}\sum_{l_1=1}^{\infty}\sum_{l_2=1}^{
\infty}a_{l_1,-1}^{\left(1\right)}a_{l_2,-1}^{\left(1\right)}\int
Y_{l,-2}^*Y_{l_1,-1}Y_{l_2,-1}d\Omega\;,\nonumber\\
a_{l,0}^{\left(2\right)}&=&\frac{1}{2}\sum_{l_1=1}^{\infty}\sum_{l_2=1}^{
\infty}a_{l_1,1}^{\left(1\right)}a_{l_2,-1}^{\left(1\right)}\int Y_{l,0}^*Y_{
l_1,1}
Y_{l_2,-1}d\Omega\;,
\label{al2}
\end{eqnarray}
and for third harmonic generation (also neglecting terms with $e^{-i\omega t}$)
%
\begin{eqnarray}
a_{l,1}^{\left(3\right)}&=&\frac{1}{2}\sum_{l_1=1}^{\infty}\sum_{l_2=1}^{
\infty}a_{l_1,-1}^{\left(1\right)}a_{l_2,-2}^{\left(2\right)}\int Y_{l,1}^*Y_{
l_1,1}
Y_{l_2,2}d\Omega\;,\nonumber\\
a_{l,-1}^{\left(3\right)}&=&\frac{1}{2}\sum_{l_1=1}^{\infty}\sum_{l_2=1}^{
\infty}a_{l_1,1}^{\left(1\right)}a_{l_2,-2}^{\left(2\right)}\int Y_{l,-1}^*Y_{
l_1,1}
Y_{l_2,-2}d\Omega\;,\nonumber\\
a_{l,3}^{\left(3\right)}&=&\frac{1}{2}\sum_{l_1=1}^{\infty}\sum_{l_2=1}^{
\infty}a_{l_1,-1}^{\left(1\right)}a_{l_2,2}^{\left(2\right)}\int Y_{l,3}^*Y_{
l_1,1}
Y_{l_2,2}d\Omega\;,\nonumber\\
a_{l,-3}^{\left(3\right)}&=&\frac{1}{2}\sum_{l_1=1}^{\infty}\sum_{l_2=1}^{
\infty}a_{l_1,-1}^{\left(1\right)}a_{l_2,2}^{\left(2\right)}\int Y_{l,-3}^*Y_{
l_1,-1}Y_{l_2,-2}d\Omega\;.
\label{al3}
\end{eqnarray}
The integrals can be expressed by the 3j-symbols and yield the coupling of the
 multipoles. Because of conservation of angular momentum, only coefficients
with $m=0,\pm2$ in SHG and $m=\pm1,\pm3$ in THG differ from zero.
%
\subsection{Higher harmonic radiated fields}
In the case of higher harmonics the electric and magnetic fields inside and
outside the sphere and the sources are matched by the boundary conditions
%
\begin{equation}
{\bf n}\cdot\left({\bf D}^{out}-{\bf D}^{in}\right)=4\pi\sigma^n\left(
\theta,\phi\right)\;,
\label{GrenzbednonlinD}
\end{equation}
and
\begin{equation}
{\bf n}\times\left({\bf E}^{out}-{\bf E}^{in}\right)=0\;.
\label{GrenzbednonlinE}
\end{equation}
As a result of spherical symmetry, only transverse magnetic waves are
generated by the oscillating surface charge. Thus the fields in the nonlinear
case are
%
\begin{equation}
{\bf
E}^i\left(\theta,\phi\right)=\sum_{l,m}\frac{m}{\left|m\right|}A_E^{\left(n
\right)}\left(l,m\right)\frac{1}{\epsilon\left(n\omega\right)k}\nabla\times f^
i_l\left(k_1r\right){\bf X}_{l,m}\left(\theta,\phi\right)\quad ,
\label{Enonlinear}
\end{equation}
where $i\equiv out$ or $i\equiv in$, respectively, and $k=n\omega/c, k_1=
\sqrt{\epsilon\left(n\omega\right)}k, f_l^{in}\left(kr\right)=j_l\left(k_1r
\right)$ and $f_l^{out}=h_l^{\left(1\right)}\left(kr\right)$. The boundary
conditions give the coefficients of the radiated field
%
\begin{equation}
A^{\left(n\right)}_E\left(l,m\right)=\frac{\frac{\partial}{\partial r}\left[
rj_l\left(k_1r\right)\right]}{\epsilon\left(n\omega\right)j_l\left(k_1r\right)
\frac{\partial}{\partial r}\left[rh_l^{\left(1\right)}\left(kr\right)\right]-
h_l^{\left(1\right)}\left(kr\right)\,\frac{\partial}{\partial r}\left[rj_l
\left(k_1r\right)\right]}\frac{\pi ka}{\sqrt{l\left(l+1\right)}}\,a_{l,m}^{
\left(n\right)}.
\label{AE(l,m)}
\end{equation}
In this case we have $k_1=\sqrt{\epsilon\left(n\omega\right)}k$.
%
\subsection{Calculation of quantities characterizing the radiation}
To study the angular dependence of the scattered field we use the quantity $
\left|E_{\phi}\left(\theta,\phi\right)\right|^2+\left|E_{\theta}\left(\theta,
\phi\right)\right|^2$ according to Born and Wolf \cite{BW}, where $E_{\theta}
\left(\theta,\phi\right)$ and $E_{\phi}\left(\theta,\phi\right)$ are the
tangential components of ${\bf E}^{sc}\left(\theta,\phi\right)$ in the linear
case and of ${\bf E}^{out}\left(\theta,\phi\right)$ in the nonlinear case.
This definition is equivalent to the absolute value of the radial part of the
Poynting vector $\left|{\bf n}\cdot\left({\bf E}\times{\bf H}\right)\right|$.
The following formulas represent $\left|E_{\phi}\left(\theta,\phi\right)
\right|^2$ and $\left|E_{\theta}\left(\theta,\phi\right)\right|^2$ in the far
field approximation. We obtain after evaluating the $m$-summation in the
linear case
%
\begin{eqnarray}
\left|E_{\phi}\left(\theta,\phi\right)\right|^2&=&\left|\sum_{l=1}^{\infty}C
\left(l\right)\left(\frac{dP_l^1\left(\cos\theta\right)}{d\theta}a_M^{sc}
\left(l,1\right)+\frac{P_l^1\left(\cos\theta\right)}{\sin\theta}a_E^{sc}\left(
l,1\right)\right)\right|^2\cdot\sin^2\phi\;,\nonumber\\
\left|E_{\theta}\left(\theta,\phi\right)\right|^2&=&\left|\sum_{l=1}^{\infty}
C\left(l\right)\left(\frac{P_l^1\left(\cos\theta\right)}{\sin\theta}a_M^{sc}
\left(l,1\right)+\frac{dP_l^1\left(\cos\theta\right)}{d\theta}a_E^{sc}\left(l,
1\right)\right)\right|^2\cdot\cos^2\phi\;,
\label{Ephithetalin}
\end{eqnarray}
for second harmonic generation
%
\begin{eqnarray}
\left|E_{\phi}\left(\theta,\phi\right)\right|^2&=&\left|\sum_{l=1}^{\infty}
\sqrt{4\pi\left(2l+1\right)}\left(\frac{dP_l^0\left(\cos\theta\right)}{
d\theta}A_E^{\left(2\right)}\left(l,0\right)\right.\right.\nonumber\\
& &\qquad\qquad\qquad\left.\left.+\frac{dP_l^2\left(\cos\theta
\right)}{d\theta}2K\left(l\right)A_E^{\left(2\right)}\left(l,2\right)\cos
\left(
2\phi\right)\right)\right|^2\;,\nonumber\\
\left|E_{\theta}\left(\theta,\phi\right)\right|^2&=&\left|\sum_{l=2}^{\infty}
\sqrt{4\pi\left(2l+1\right)}\frac{dP_l^2\left(\cos\theta\right)}{d\theta}
A_E^{\left(2\right)}\left(l,2\right)\right|^2\cdot\sin^2\left(2\phi\right)\;,
\label{EphithetaSHG}
\end{eqnarray}
and third harmonic generation
%
\begin{eqnarray}
\left|E_{\phi}\left(\theta,\phi\right)\right|^2&=&\left|\sum_{l=1}^{\infty}
\sqrt{4\pi\left(2l+1\right)}\left(\frac{P_l^1\left(\cos\theta\right)}{\sin
\theta}A_E^{\left(3\right)}\left(l,1\right)\sin\phi\right.\right.\nonumber\\
& &\qquad\qquad\qquad\left.\left.+\frac{P_l^3\left(\cos\theta
\right)}{\sin\theta}K\left(l\right)A_E^{\left(3\right)}\left(l,3\right)\sin
\left(3\phi\right)\right)\right|^2\;,\nonumber\\
\left|E_{\theta}\left(\theta,\phi\right)\right|^2&=&\left|\sum_{l=1}^{\infty}
\sqrt{4\pi\left(2l+1\right)}\left(\frac{dP_l^1\left(\cos\theta\right)}{d
\theta}\right.\right.
A_E^{\left(3\right)}\left(l,1\right)\cos\phi\nonumber\\
& &\qquad\qquad\qquad+\left.\left.\frac{dP_l^3\left(\cos\theta\right)
}{d\theta}K\left(l\right)A_E^{\left(3\right)}\left(l,3\right)\cos\left(3\phi
\right)\right)\right|^2\;,
\label{EphithetaTHG}
\end{eqnarray}
where $K\left(l\right)$ are $l$- and $m$-dependent factors.

Note that the Legendre polynomials $P_l^m$ with $m=0,2$ are identically zero
for $\theta=0,\pi$, since there is no direct scattering in forward nor
backward direction in the second harmonic case. Furthermore the $\phi$-
dependence of the linear scattering and THG is described by the interval
($0,\pi$) and by $\left(0,\frac{\pi}{2}\right)$ in SHG according to the
symmetries of the sine and cosine terms.

We will plot the degree of polarization defined as
%
\begin{equation}
P\left(\theta\right)=\frac{I_{\parallel}-I_{\perp}}{I_{
\parallel}+I_{\perp}}\;,
\label{polarisation}
\end{equation}
with
%
\begin{equation}
I_{\parallel}=\left|E_\theta\left(\theta,\phi=\frac{\pi}{2}\right)\right|^2+
\left|E_\phi\left(\theta,\phi=\frac{\pi}{2}\right)\right|^2\;,
\label{Ipar}
\end{equation}
and
%
\begin{equation}
I_{\perp}=\left|E_\theta\left(\theta,\phi=0\right)\right|^2+\left|E_\phi\left(
\theta,\phi=0\right)\right|^2\;.
\label{Isen}
\end{equation}
To measure the asymmetry of forward and backward scattering in the $Mie$-range
 we introduce the quantity
%
\begin{equation}
R=\frac{I_{forw}-I_{back}}{I_{forw}+I_{back}}\;,
\label{VorwRueck}
\end{equation}
which we call the ``degree of Mie-asymmetry''. In the linear case $I_{forw}$
and $I_{back}$ are the scattering intensities taken at $\theta=0$ and $\theta=
\pi$ respectively. As these quantities are identically zero in SHG, we use for
 $I_{forw}$ and $I_{back}$ the maxima of the scattering intensities along the
direction of propagation of the incident wave for $\phi=0,\frac{\pi}{4},\frac{
\pi}{2}$ and $\theta$ covering the interval $\left(0,\pi\right)$. In THG the
angular dependence fo the radiated intensities is more complicated compared to
the linear case and we take the maximum for $\phi=0$ and $\theta$ ranging
from 0 to $\pi$.

Finally, we calculate the angle-integrated scattered intensities. We obtain
in the linear case
\begin{equation}
Q_{sc}^{\left(1\right)}=\frac{1}{\pi\left(ka\right)^2}\sum_{l,m}\frac{2l+1}{l
\left(l+1\right)}
\left(\left|a^{sc}_E\left(l,m\right)\right|^2+\left|a_M^{sc}\left(l,m\right)
\right|^2\right)\quad ,
\label{Qsclin}
\end{equation}
and for the $n$-th harmonic
%
\begin{equation}
Q^{\left(n\right)}_{sc}=\frac{1}{\pi\left(ka\right)^2}\sum_{l,m}\left|A_E^{
\left(n\right)}\left(l,m\right)\right|^2\quad.
\label{Qscnonlin}
\end{equation}
Here $Q^{\left(n\right)}_{sc}$ is in units of the geometric cross section of
the sphere $\pi a^2$.
In this formulation the optical theorem, which links the extinction efficiency
 to the normalized scattering amplitude in forward direction, has the form
%
\begin{equation}
Q_{ext}=\frac{2}{\pi\left(ka\right)^2}\cdot\left(\left|E_{\theta}\left(\theta=0\right)\right|^2+\left|E_{\phi}\left(\theta=0\right)\right|^2\right)\;,
\label{optictheo}
\end{equation}
with
%
\begin{equation}
Q_{ext}=\frac{1}{\pi\left(ka\right)^2}\sum_{l,m}\left(2l+1\right)\left\{Re
\left[a_E^{sc}\left(l,m\right)\right]+Re\left[a_M^{sc}\left(l,m\right)\right]
\right\}
\label{Qext}.
\end{equation}
\newpage
\section{Numerical results}
We present numerical results for the {\em angular dependence} of the radiated
intensities obtained using Eqs. (\ref{Ephithetalin})-(\ref{EphithetaTHG}). The
 {\em degree of polarization} characterizing the radiation, for example its
angular dependence. The {\em degree of Mie-asymmetry} $R$ calculated using
Eq.(\ref{VorwRueck}) gives us the strength of the asymmetry along the
direction of propagation according to the {\em Mie-effect}. The {\em
integrated intensities} give us the resonances as a funtion of cluster size.
In all cases we compare the linear results with the results of second and
third harmonic generation as a function of the size parameter $ka$ and the
material-properties, referring to iron and nickel at a fixed optical
wavelength of 617~nm.
To check the numerical accuracy we compare the linear results with those of
ref.\cite{BH}. We find excellent agreement. Furthermore, we check the optical
theorem. By determining $\Delta Q\equiv\left|Q_{ext} - 2\cdot\left(\left|E_{
\theta}\right|^2+\left|E_{\phi}\right|^2\right)_{\theta=0}/\left[\pi\left(ka
\right)^2\right]\right|$ as a function of $ka$ we find that this is satisfied
to an accuarcy better than $10^{-12}$.\\
As input parameter we take the complex refractive indices measured by Johnson
and Christy \cite{JohnChrist}. In the linear case, we also use for comparison
with other calculations the refractive index of water droplets, as given by
Bohren and Huffman~\cite{BH}. These values are listed in Tab. I. The
refractive index is constant in all figures unless something else is
specified. Thus varying the size parameter $ka$ means varying the size of the
sphere.
%
\subsection{Angular dependence of the intensities}
First, we show polar plots of $\left|E_{\phi}\right|^2+\left|E_{\theta}\right|
^2$ (see Figs. \ref{lin3d} - \ref{shg3d}). In general, the shape of the plots
is governed by the values of the coefficients in the series expansions Eqs.
(\ref{Ephithetalin})-(\ref{EphithetaTHG}). It is well known \cite{Jackson}
that in the linear series (\ref{Ephithetalin}) only terms with $l\leq ka$
contribute significantly. For $l>ka$ the terms decrease very rapidly, whereas
for $l\ll ka$ they have comparable magnitudes. We restrict our calculation
to $l<\left[{\rm Max}\left(\tilde{n}\cdot ka,\tilde{n}'\cdot ka\right)+15
\right]$ with a maximum value of $l$ of 50 where $\tilde{n}$ and $\tilde{n}'$
are the real and imaginary parts of the complex refractive index. This gives
satisfactory convergence of the series up to $ka<10$ in all harmonics. To
qualify the angular dependence the range of size parameters up to 5 is
sufficient. In this range the $l$-values of the dominating terms are a little
smaller than $ka$ and no terms with $l\ll ka$ exist. Terms with $l\leq ka$ can
 be very different from each other in contrast to terms with $l\ll ka$ which
have comparable magnitudes. Thus, a pronounced transition range from pure
dipole-scattering to pure {\em Mie}-scattering exists. Of course we cannot
reach the transition from {\em Mie}-scattering to the optical limit of
reflection, due to the numerical limit of a maximum value of $l$ of 50.

The geometry of the scattering is specified in the inset of Fig. \ref{lin3d}
with the direction of propagation of the incident wave being parallel to the
positive y-axis and polarization along the positive z-axis. Fig.~\ref{lin3d}~
(a) shows {\em Rayleigh}-scattering according to the dipole-term with $l=1$.
The characteristic $\cos^2\theta$-dependence appears along the x-z-plane. The
other plots for linear optics (Figs. \ref{lin3d}) show the well known results
\cite{Mie,BH,BW}. For a value of $ka=1$ asymmetry of forward and backward
 scattering appears according to the {\em Mie-effect}. The ratio of forward
to backward scattering $I_{forw}/I_{back}$ increases strongly with increasing
size parameter $ka$ beginning at a value of $ka\approx 1$. ``New'' maxima grow
 out in the backward direction and move to the forward direction with
increasing $ka$. The $\phi$-dependence is not as striking and not as
complicated as the $\theta$-dependence since we have a superposition of the
form $A\left(\theta\right)\cdot\cos^2\phi+B\left(\theta\right)\cdot\sin^2\phi$
, the scattering-behavior is dominated by the strong increase in forward
scattering described by the $\theta$-dependence. The fact that the $\phi$-
dependence is described by the intervall $\left(0,\pi\right)$ is most
important for the linear case and harmonics with odd order and differs from
SHG and harmonics or even order where the $\phi$-dependence is fully described
 by the intervall ($0,\frac{\pi}{2}$).

The polar-plots in the case of THG are quite similar to the linear case up to
values of $ka\approx 2$. The differences between Figs. \ref{lin3d} (b) and
\ref{thg3d} (b) with $ka=1$ reflect the stronger increase in the ratio of
forward to backward scattering in THG with increasing $ka$. The plots in
Figs. \ref{lin3d} (c) and \ref{thg3d} (c) with $ka=2$ are very similar apart
from one more maximum appearing in the third harmonic case for $\theta\approx\
frac{\pi}{2}$. For $ka=5$ the intensities parallel to the direction of
polarization are much larger in third harmonic than in the linear case. Note,
the different scales of the axes for different $ka$. The terms for $m=3$ in
THG are negligble compared to the terms with $m=1$. So the differences in the
magnitudes of the linear and third harmonic intensities are caused by the
coefficients $a^{sc}_E\left(l,1\right), a^{sc}_M\left(l,1\right)$ and $A^{
\left(3\right)}_E\left(l,1\right)$ only.

The angular dependence of the intensities in second harmonic is very different
 from the linear and third harmonic cases. The disappearing direct forward and
 backward intensities and the $\cos^2\left(2\phi\right)$ and $\sin^2\left(2
\phi\right)$ behavior produce the club-shaped structure, which is shown in
Figure \ref{shg3d}. But the main features of the linear and third harmonic
plots appear also in second harmonic. The plots become asymetrical with
respect to $\theta$ in the range of $ka\approx1$. The ratio of forward to
backward scattering increases with $ka$ and is between the linear and third
harmonic one. Values of the forward to backward scattering ratio with
different $ka$-values are listed in Tab. II. In general harmonics with even
order will show an angular dependence like SHG, because the Legendre
polynomials of $m\neq 1$ vanish at $\theta=0,\pi$ and $\phi$ will appear in
the cosine and sine-terms in connection with $n=0,2,.,2p$ ($p$ integer).
Analogous harmonics with odd order will behave similar to the linear case.
\subsection{Polarization}
To learn more about the angular dependence, especially the $ka$-dependence of
the intensity maxima,  we calculate the degree of polarization using
Eq.~(\ref{polarisation}). First, we plot the  polarization in the case of
{\em Rayleigh}-scattering (Fig. \ref{linpol} (a)). For convenience and in
agreement with the preceding section each of the figures in Fig.~\ref{linpol}
contain the linear and third harmonic curves. The plots of {\em Rayleigh}-
scattering (Fig. \ref{linpol} (a)) are identical. With increasing size
parameter the maximum of the polarization moves to smaller angles (the forward
 direction) in agreement with the intensity maxima. In third harmonic
generation they move faster than in the linear case (Fig. \ref{linpol} (b)).
For $ka=2$ (Fig. \ref{linpol} (c)) there is one more maximum in THG in
agreement with the Figs. \ref{lin3d} (c) and \ref{thg3d} (c). Increasing the
size parameter up to 5 destroys the correlation between the peaks in the
polarization plots and the polar plots in THG. For example the ``double''
peaks in the third harmonic plot with $ka=5$ cannot be identified with
``double'' peaks in the polar plots but they reflect that the intensities at
$\phi=0$ are comparable to those at $\phi=\frac{\pi}{2}$ in contradiction to
the linear case. In the linear case the polarization even for values higher
than $ka=5$ reproduces the position of the peaks in the polar plots and is
mainly perpendicular ($P\left(\theta\right)>0$). This is an artifact of the
chosen refractive index and is not characteristic for  linear {\em Mie}-
scattering in general. For imaginary parts of the refractive index close to
zero, the polarization is mainly perpendicular.

The polarization plots in second harmonic generation reflect the different
shape of the polar plots. Since the $P\left(\theta=0\right)$ and $P\left(
\theta=\pi\right)$ values do not exist and the limits $\lim_{\theta\rightarrow
 0}P\left(\theta\right)$ and $\lim_{\theta\rightarrow\pi}P\left(\theta\right)$
 are different, the plot in the {\em Rayleigh}-range is asymmetric.
The decrease of $\left|P\right|$ up to zero with increasing $ka$ is
characteristic for SHG. The square of the absolute value of the $\theta$-
component of the electric field $\left|E_{\theta}\right|^2$ is identically
zero for $\phi=0,\pi$~. Thus $P\left(\theta\right)$ is a measure of the
importance of the ($m=2$)-term in $\left|E_{\phi}\left(\theta,\phi\right)
\right|^2$ with respect to the ($m=0$)-term. Thus the ($m=0$)-term can be
neglected with increasing $ka$. Only for very small $ka$, the structure in
the polarization plots correlates with the shape of the polar plots. The first
 ``new'' maximum appears as zero in the polarization. For larger $ka$,
however, there is no correlation any more.
\newpage
\subsection{Forward vs. backward scattering}
By computing $R$ defined in Eq. (\ref{VorwRueck}) which is a measure of the
difference between the forward and backward intensities as a function of the
size parameter $ka$ and the real or imaginary part of the refractive index
$N=\tilde{n}+i\cdot\tilde{n}'$, we want to study the development of the
asymmetry of forward to backward intensities. Of particular interest will be
enhanced backward scattering.

Figure \ref{linvr} (a) shows the increase of forward scattering with the size
of the sphere in linear scattering. The oscillations correlate with intensity
maxima in the backward direction resulting from maxima of the coefficients,
see Probert-Jones~\cite{ProbJones}. Enhanced backward scattering appears only
in the small range of $0.5<ka<1$ in the case of iron. The curve for water
droplets shows no enhanced backscattering, but the overall behavior is the
same as for metals.
Varying the imaginary part of the refractive index diminishes the forward
scattering. In the case of $ka=1$ and $\tilde{n}'>5$ an enhancement of the
backward scattering occurs. The $\tilde{n}$ dependence of $R$ is similar.

In the higher harmonic case the {\em Mie-effect} is strongly enhanced. The
limit of one is obtained earlier than in the linear case. The higher the
order, the stronger is the enhancement of the forward scattering. Regions of
enhanced backward scattering are hard to find. In the case of second harmonic
generation, they exist only for small $\tilde{n}$ or $\tilde{n}'$ around 5 and
 small $ka$ whereas we could not find enhanced backward scattering for metal-
clusters ($\tilde{n}'\gg 0$) in third harmonic generation so far.
\newpage
\subsection{Integrated intensities}
{\em Mie}-Resonances appear with increasing radius $a$ of the sphere or
absolute value of the wave vector $k$ of the incident light. In this section
we will only deal with the size-dependent resonances.

In Fig. \ref{linQsc133.0} we show the scattering efficiency $Q_{sc}$ of water
droplets as a function of the size parameter. The main features are the
interference structure built up by interferences between the incident wave and
 forward scattered light and the ripple structure reflecting resonant surface
modes. Furthermore, the results suggest the optical paradoxon $\lim_{ka
\rightarrow\infty}Q_{ext}\left(ka\right)=2$. The ripple structure correlates
with resonances in the coefficients $a_E^{sc}\left(l,m\right)$ and $a_M^{sc}
\left(l,m\right)$. They are resonant if their imaginary part is zero. Fig.
\ref{kres} shows the first resonance of $a_E^{sc}\left(13,1\right)$ for a real
 refractive index with $\tilde{n}=1.5$ and $\tilde{n}'=0$. The resonance of
$a_M^{sc}\left(13,1\right)$ occurs for $ka\approx 11$ (see Ch\'ylek
\cite{Chylek}). For large size parameters the distance of the ripples can be
expressed directly by the refractive index. Finite imaginary parts of the
refractive index would damp the resonance. Then the values of the coefficients
 at the resonant point would be smaller than 1. Even if we take $\tilde{n}'
\approx 0.1$ the ripple structure in Fig. \ref{kres} does not appear.
Accordingly, there is no ripple structure in the scattering efficency of Fe
and Ni as a function of size as shown in Fig. \ref{QscKA} (a).

In higher harmonics no structure correlated with resonances of the coefficient
 can be detected even for vanishing imaginary part of the refractive index up
to a numerical accuracy of $10^{-10}$. Each coefficient $a^{\left(n\right)}_{
l,m}$ is a combination of all linear electric multipole coefficients
determined by Eqs. (\ref{al2}) and (\ref{al3}), respectively. Correspondingly,
 the size dependence of the coefficients $A_E^{\left(n\right)}\left(l,m\right)
$ has many small peaks (Fig. \ref{kres} (b) and (c)). In contrast, the size
resonances caused by multipole-combinations, and known as interference
structure in linear scattering, are in the case of metals more pronounced in
SHG (compare Fig. \ref{QscKA} (a) and (b)). In THG (Fig. \ref{QscKA} (c)) the
first peak is even more dominant and the resonances are visible only for
enhanced resolution. In both cases, the position of the resonances has a weak
dependence on the refractive index, but the number of peaks in SHG is twice
the peak-number in THG if $0<ka<10$. Decreasing the imaginary part of the
refractive index down to zero changes the behavior drastically. Instead of a
dominating first peak the scattering efficiency now shows an continuous
increase for $0<ka<20$ with oscillations stronger than in the case of $\tilde{
n}'>1$. In all three cases the absolute values grow with the absolut value of
the refractive index if metals are considered.
\newpage
\section{Summary}
By extending the classical model introduced by \"Ostling {\em et al.} \cite{
Stampf} we calculated the angular dependence of the second and third harmonic
intensities radiated by spherical metal clusters. Therein the n-th power of
the surface charge density induced by linear polarized light is used as the
source of the fields radiated in higher harmonics \cite{add}. The source
represents the discontinuity of the electrical displacement at the surface of
the sphere.

We find that the forward Mie-scattering is in the nonlinear case even more
strongly enhanced (see Table II). The nonlinear optical response yields a
stronger size sensitivity. Higher multipoles contribute already at smaller
size parameters $ka$. For example, we find that the light intensities
perpendicular to the forward direction divided by the light intensities in
forward direction is much larger in the third harmonic than in the linear
case. Since the values of the Legendre polynomials with azimuthal quantum
numbers $m=0,\pm2$ vanish for $\theta=0,\pi$ and due to the $\phi$-dependence
of the form $A\cdot\cos^2\left(2\phi\right)+B\cdot\sin^2\left(2\phi\right)$,
see Eq. (\ref{EphithetaSHG}), the angular dependence of the second harmonic
intensities is very different from the linear and third harmonic
distributions. Especially, direct forward and backward scattering vanishes.
In contrast, the linear {\em Mie}-scattering results (terms with $m=\pm1$
only) and THG (terms with $m=\pm1,\pm3$) are similar. In particular the terms
with $m=\pm3$ in THG and $m=0$ in SHG can be neglected, since their absolute
values are much smaller than the other contributions. The more pronounced {\em
 Mie-effect} prevents enhanced backward scattering in higher harmonics (Figs.
 \ref{nhgvrKA}).

In order to compare with our theory an experimental investigation of the
nonlinear {\em Mie}-scattering would be interesting. {\em Mie}-resonances
play an important role in the field of photonic bandstructure in high
refractive materials (see John \cite{John}), but they are studied so far in
the linear case only. Also applications of the theory to the study of
fullerenes and problems in biophysics (detection and growth modes of tumor
cells) would be interesting.

We will extend the theory to ellipsoidal objects to solve the classical
problem of a one to one correspondence of scattering profile and particle
shape. Furthermore, the expected curvature sensitivity of the nonlinear
optical response even to particles with sizes much smaller than the wavelength
, should be detectable in the nonlinear scattering.
Thus, the higher harmonics are a particularly useful probe for detecting small
 particle sizes and shapes.
\newpage
%
\begin{figure}
\caption{The linear extinction $Q_{ext}\left(ka\right)$ as a function of the
size parameter $ka$ and varying the radius of the sphere for water droplets
with complex refractive index $N=1.33+i\cdot 10^{-7}$.}
\label{linQsc133.0}
\vspace{0.5cm}
\caption{Difference $\Delta Q=Q_{ext}-2\cdot\left(\left|E_{\theta}\right|^2+
\left|E_{\phi}\right|^2\right)_{\theta=0}/\left[\pi\left(ka\right)^2\right]$
of the extinction efficiency and the modified forward scattering amplitude, as
 a function of the size parameter $ka$. This illustrates how our numerical
calculations fulfill the optical theorem.}
\label{opttheo}
\vspace{0.5cm}
\caption{Polar plots of the linear scattered intensities for Fe: (a) scattered
 linear intensity $I_{\omega}\left(\theta,\phi\right)$ in units of $10^{-12}$
with $ka=0.001$, (b) $I_{\omega}\left(\theta,\phi\right)$ in units of
$10^{-4}$ with $ka=1$, (c) $I_{\omega}\left(\theta,\phi\right)$ in units of 1
with $ka=2$, and (d) $I_{\omega}\left(\theta,\phi\right)$ with $ka=5$ in units
 of 1. The inset shows the scattering geometry. The direction of the incident
light is determined by $\theta=0$ with polarization along the axis defined by
$\phi=0$ and $\theta=\pi/2$.}
\label{lin3d}
\vspace{0.5cm}
\caption{Polar plots of the third harmonic (THG) intensities for Fe: (a)
scattered $3\omega$-intensity $I_{3\omega}\left(\theta,\phi\right)$ in umits
of $10^{-15}$ with $ka=0.001$, (b) $I_{3\omega}\left(\theta,\phi\right)$ in
units of 1 with $ka=1$, (c) $I_{3\omega}\left(\theta,\phi\right)$ in units of
$10^{-5}$ with $ka=2$, and (d) $I_{\omega}\left(\theta,\phi\right)$ in units
of $10^{-5}$ with $ka=5$.}
\label{thg3d}
\vspace{0.5cm}
\caption{Polar plots of the second harmonic (SHG) intensities for Fe: (a)
scattered $2\omega$-intensity $I_{2\omega}\left(\theta,\phi\right)$ in units
of $10^{-18}$ with $ka=0.001$, (b) $I_{2\omega}\left(\theta,\phi\right)$ in
units of 1 with $ka=1$, (c) $I_{2\omega}\left(\theta,\phi\right)$ in units of
$10^2$ with $ka=2$ and, (d) $I_{\omega}\left(\theta,\phi\right)$ in units of
$10^6$ with $ka=5$.}
\label{shg3d}
\vspace{0.5cm}
\caption{Degree of polarization $P$ for the linear and third harmonic case as
a function of the polar angle $\theta$ (a) for $ka=0.001$, (b) for $ka=1.$,
(c) for $ka=2$, and (d) for $ka=5$.}
\label{linpol}
\vspace{0.5cm}
\caption{Degree of polarization $P$ for the second harmonic case as a function
 of the polar angle $\theta$ (a) for $ka=0.001$, (b) for $ka=1$, (c) for
$ka=2$, and (d) for $ka=5$.}
\label{shgpol}
\vspace{0.5cm}
\caption{Degree of {\em Mie-asymmetry} $R$ (defined as $R=\left(I_{forw}-I_{
back}\right)/\left(I_{forw}+I_{back}\right)$, see Eq.(23)) in the linear case
(a) as a function of the size parameter $ka$ for refractive indices referring
to Fe, Ni and water droplets, (b) as a function of the real part of the
refractive index $\tilde{n}$, and (c) as a function of the imaginary part of
the refractive index $\tilde{n}'$ for size paramters $ka=1$ and $ka=5$ using
values of $\tilde{n},\tilde{n}'$ for Fe and Ni.}
\label{linvr}
\vspace{0.5cm}
\caption{Degree of {\em Mie-asymmetry} $R$ as a function of the size parameter
 $ka$ for (a) SHG, and (b) THG with the refractive indices referring to Fe and
 Ni, respectively.}
\label{nhgvrKA}
\vspace{0.5cm}
\caption{Degree of {\em Mie-asymmetry} $R$ as a function of the real part of
the refractive index $\tilde{n}$. For $ka=1$ and $ka=5$ for (a) SHG, and (b)
THG. $\tilde{n}'$ corresponds to Fe and Ni.}
\label{nhgvrnIIr}
\vspace{0.5cm}
\caption{Degree of {\em Mie-asymmetry} $R$ as a function of the imaginary part
 of the refractive index $\tilde{n}'$. For $ka=1$ and $ka=5$ for (a) SHG, and
(b) THG. $\tilde{n}$ corresponds to Fe and Ni.}
\label{nhgvrnIIi}
\vspace{0.5cm}
\caption{(a) Real and imaginary part of the expansion coefficient of the
scattered linear fields $a_E^{sc}\left(l=13,m=1\right)$ as a function of size
parameter $ka$ and for a real refractive index $\tilde{n}=1.5$. The first peak
 represents the lowest resonance of the multipole mode with $l=13$. The real
and imaginary parts of the expansion coefficients of the SHG and THG fields $
A_{8,2}^{\left(2\right)}$ and $A_{8,1}^{\left(3\right)}$, respectively, as a
function of the size parameter $ka$ are shown in Figs. (b) and (c). The
results in (b) and (c) show that the resonance peaks of the linear
coefficients contribute to the higher harmonic coefficients: The maxima and
minima of $A^{\left(2\right)}$ and $A^{\left(3\right)}$ are essentially
determined by the resonances of the linear coefficients $a_E^{sc}$.}
\label{kres}
\vspace{0.5cm}
\caption{Scattering efficiencies $Q_{sc}$ as a function of size parameter $ka$
 for (a) linear scattering (b) SHG, and (c) THG in the case of Fe and Ni,
respectively.}
\label{QscKA}
\end{figure}
\begin{table}
\caption{Ratio of forward to backward intensities as a function of $ka$ in the
 linear case, SHG and THG.}
\label{forwbacktab}
\end{table}
\begin{center}
\begin{tabular}{|c||c|c|c|}
\hline
    & \multicolumn{3}{c|}{$I_{forw}/I_{back}$}\\ \hline
$ka$& lin & SHG & THG \\ \hline\hline
1   & 1.4 & 2.7 & 15  \\
2.1 & 8.7 & 10  & 113 \\
4.8 & 64  & 136 & 206 \\ \hline
\end{tabular}
\end{center}
\begin{table}
\caption{Complex refractive indices $N=\tilde{n}+i\tilde{n}'$ for iron and
nickel used in this work at $\omega, 2\omega$ and $3\omega$. The wavelength is
 $\lambda=617~nm$ and $\omega=2\pi/\lambda\cdot c$.}
\label{nktab}
\end{table}
\begin{center}
\begin{tabular}{|c||c|c|c|c|}
\hline
          & \multicolumn{2}{c|}{iron} & \multicolumn{2}{c|}{nickel} \\ \hline
          & $\tilde{n}$& $\tilde{n}'$ & $\tilde{n}$  & $\tilde{n}'$ \\ \hline
                                                                       \hline
 $\omega$ & 2.88       & 3.05         & 1.99         & 4.02         \\
$2\omega$ & 1.69       & 2.06         & 2.01         & 2.18         \\
$3\omega$ & 1.49       & 1.41         & 1.29         & 1.89         \\ \hline
\end{tabular}
\end{center}


\newpage
\begin{thebibliography}{99}
\bibitem{Mie}G. Mie, Ann. Phys. (Leipzig) {\bf 25}, 377 (1908)
\bibitem{BH}C. F. Bohren and D. R. Huffman: {\em Absorption and Scattering of
Light by Small Particles}, (Wiley, New York, 1983)
\bibitem{Martin}S. S. Martinos, Phys. Rev. B {\bf 31}, 2029 (1985)
\bibitem{Stampf}D. \"Ostling, P. Stampfli and K. H. Bennemann, Z. Phys. D {\bf
 28}, 169-175 (1993)
\bibitem{Jackson}J. D. Jackson: {\em Classical Electrodynamics} (Wiley, New
York, 1975)
\bibitem{BW}M. Born and E. Wolf: {\em Principles of Optics}, (Pergamon Press,
Oxford, 1975)
\bibitem{JohnChrist}P. B. Johnson, R. W. Christy, Phys. Rev. B {\bf 9}, 5056-
5070 (1974)
\bibitem{ProbJones}J. R. Probert-Jones, J. Opt. Soc. Am. A, {\bf 1} 8, 822
(1984)
\bibitem{Chylek}P. Ch\'ylek, J. Opt. Soc. Am., {\bf 66} 3, 285-287 (1976)
\bibitem{add}In addition the following approximations are made in ref.~\cite{
Stampf}: (a) The nonlinear susceptibilities are taken to be frequency
independent, $\chi^{\left(n\right)}\left(\omega\right)=$const., (b) the
surface charge is assumed to be $\sigma\left(n\omega\right)=c\cdot\sigma^n
\left(\omega\right)$, and (c) the proportionality-factor $c$ has not been
determined by \"Ostling {\em et al.} since they focus on relative intensities.
 It would be of interest to compare these approximations with a detailed
microscopic theory of the nonlinear susceptibilities $\chi^{\left(n\right)}
\left(\omega\right)$.
\bibitem{John}S. John, {\em Localization of Light}, Physics Today, May 1991
\end{thebibliography}
\end{document}